# New Particle Identification Approach with Convolutional Neural Networks in GAPS

By Masahiro Yamatani,[1] Yusuke Nakagami,[2] Hideyuki Fuke,[1] Akiko Kawachi,[3]
Masayoshi Kozai,[4] Yuki Shimizu,[5] and Tetsuya Yoshida[1]

[1] *Institute of Space and Astronautical Science, JAXA, Sagamihara, Japan*
[2] *Graduate School of Science and Engineering, Aoyama Gakuin University, Sagamihara, Japan*
[3] *School of Science, Tokai University, Hiratsuka, Japan*
[4] *Joint Support-Center for Data Science Research, Research Organization of Information and Systems, Tachikawa, Japan*
[5] *Faculty of Engineering Department of Physics, Kanagawa University, Yokohama, Japan*

The General Antiparticle Spectrometer (GAPS) is a balloon-borne experiment that aims to measure low-energy cosmic-ray antiparticles. GAPS has developed a new antiparticle identification technique based on exotic atom formation caused by incident particles, which is achieved by ten layers of Si(Li) detector tracker in GAPS. The conventional analysis uses the physical quantities of the reconstructed incident and secondary particles. In parallel with this, we have developed a complementary approach based on deep neural networks. This paper presents a new convolutional neural network (CNN) technique. A three-dimensional CNN takes energy depositions as three-dimensional inputs and learns to identify their positional/energy correlations. The combination of the physical quantities and the CNN technique is also investigated. The findings show that the new technique outperforms existing machine learning-based methods in particle identification.

## 1. Introduction

The existence of dark matter (DM) is confirmed by various measurements, including the rotational velocity of spiral galaxies, gravitational lensing, and the universe's large-scale structure. However, its true nature remains unknown.[1] Several observations indicate that weakly interacting massive particles (WIMPs) are strong candidates for DM. WIMPs are very attractive because they can solve several problems of the Standard Model of particle physics (hierarchy problem, gauge coupling unification, etc.). To cover the WIMP's vast parameter space, performing DM searches through complementary approaches such as direct, collider, and indirect detections are critical.

The indirect detection of DM using antideuterons can be a new probe for exploring WIMP parameters.[2,3] Antideuterons can be produced using the coalescence of standard model particles originating from WIMP annihilation or decay. Although secondary or tertiary antideuterons can be created, their origin is cosmic-ray hadronic interaction, where antideuteron production is kinematically suppressed in the sub-GeV low-energy region.[4,5] Thus, the expected yield is several orders of magnitude lower than DM-originated antideuterons. Therefore, detecting a single sub-GeV antideuteron candidate can provide secure evidence of a new origin. The predicted antideuteron flux corresponding to WIMP parameters, indicated by AMS–02 antiproton excess (interpreted as annihilation into purely $b\bar{b}$),[6,7] is shown as a cyan band in Fig. 1., as well as that from the secondary (blue) and tertiary (light purple) backgrounds.[3] Each prediction band's width The width corresponds to uncertainty in coalescence momenta.

## 2. GAPS Project

The General Antiparticle Spectrometer (GAPS) project is a

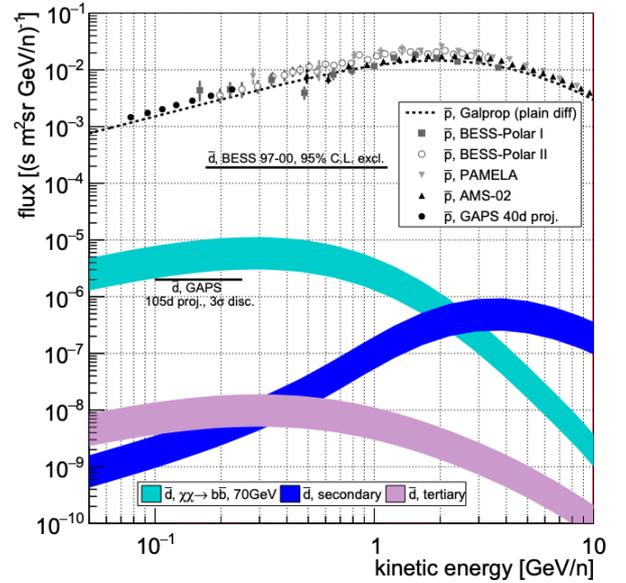

Fig. 1. Predicted GAPS sensitivity to antiprotons (upper points) and antideuterons (lower bands) as a function of kinetic energy per nucleon. The anticipated GAPS antiproton measurement after one flight is shown compared with the GALPROP plain diffusion prediction and current spectra from AMS-02, BESS-Polar I/II, and PAMELA. The anticipated GAPS 3σ antideuteron discovery sensitivity by three long-duration balloon flights is also shown, and it reaches the predicted antideuteron flux corresponding to WIMP parameters indicated by AMS-02 antiproton excess.[3]

balloon-borne experiment contributing to DM physics by searching cosmic-ray antiparticles.[3,8] The main objective of GAPS is to search for undiscovered cosmic-ray antideuterons in the low-energy region below 0.25 GeV/n. For a highly sensitive search, GAPS plans to conduct several Antarctic balloon experiments using NASA's Long Duration Balloons

(LDBs). Fig. 1 depicts the expected antideuteron sensitivity of GAPS for three LDB flights. The sensitivity could reach the antideuteron flux predicted using reasonable DM models.[3]

GAPS can detect more antiprotons in the low energy region (upper plots in Fig. 1)[9] than previous experiments such as BESS-Polar and PAMELA by one order of magnitude. This precise measurement in the low energy region provides new phase space to investigate light DM models such as light neutralinos,[10] gravitinos,[11] and LZPs.[12] Antihelium, another new probe for DM physics, is also an antiparticle to which GAPS is highly sensitive.[13] However, this paper does not cover searches for antiprotons and antihelium.

### 2.1. Detection concept

High grasping power and high discrimination power are necessary to observe DM-induced antiparticles. For instance, the typical fluxes of cosmic-ray proton and antiproton backgrounds are approximately $10^{11}$ and $10^4$, respectively, greater than the antideuteron flux shown in Fig. 1. GAPS introduced a unique method using exotic atom's de-excitation sequence to achieve strong discrimination power in the presence of such background events.[14,15]

Figure 2 presents an overview of the GAPS detector. A time-of-flight (TOF) system composed of plastic scintillators covers the central tracker, consisting of more than 1,000 lithium-drifted silicon Si(Li) detectors.

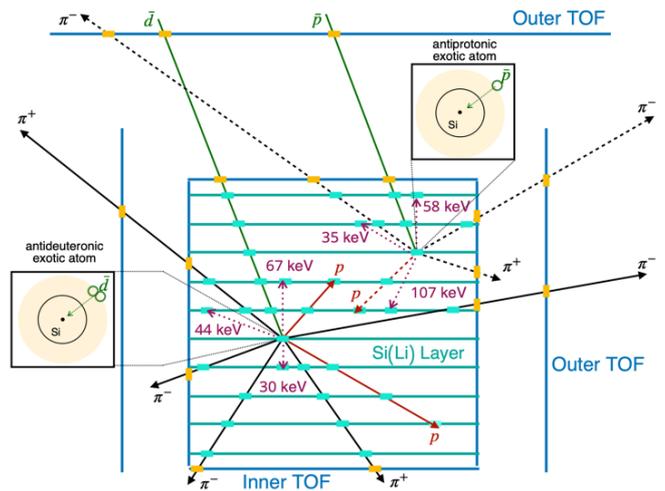

Fig. 3. Schematic of GAPS antiparticle identification. Antiparticles decelerate and stop in the Si(Li) target to form exotic atoms. De-excitation of these exotic atoms emits characteristic X-rays, followed by nuclear annihilation, which releases pions and protons. This technique identifies antiprotons and antideuterons in cosmic rays with high detection efficiency and sufficient rejection against the background.[16]

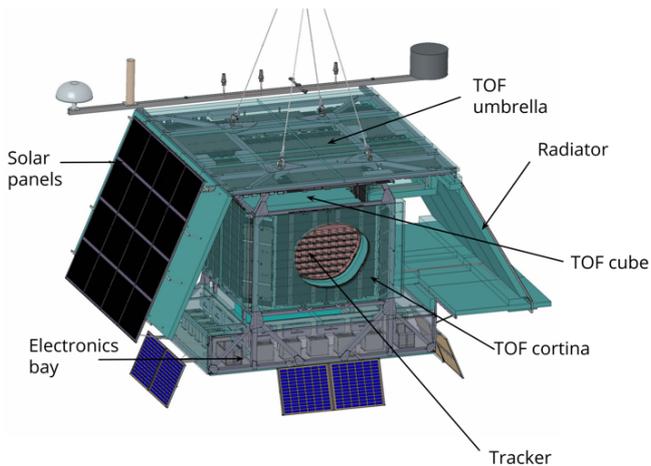

Fig. 2. Conceptual diagram of the GAPS instrument configuration. The tracker in the center of the instrument has 1440 Si(Li) detectors arranged over ten layers. The tracker is surrounded by inner and outer TOF plastic scintillation counters.

Figure 3 shows an overview of the GAPS antiparticle identification technique using the exotic atom formation.[16] When an antiparticle enters the instrument, it passes through the TOF counter and the Si(Li) tracker sequentially, depositing energy at each hit. Then, the incident particle stops in the Si(Li) tracker and is trapped in the nucleus to form an excited exotic atom. The characteristic X-ray energies and the secondary particle multiplicity are strictly determined by the physics of the exotic atoms and can be used for antiparticle identification. The detector can also obtain pion and proton trajectories using hit information. They overcome the technical limitations of conventional magnetic spectrometers requiring heavy magnets, enabling the development of instruments with large grasping power and low-energy ranges. A test beam at KEK demonstrated this principle.[17,18]

### 2.2. Instrument design

The central detectors for particle tracking comprise ten layers of 1440 Si(Li) detectors[19-22] with 10 cm spacing vertically, as shown in Fig. 2. The volume is 1.6m × 1.6m × 1m. A 2.5-mm- thick Si(Li) wafer has eight strips contained in 4-inch diameter. The Si(Li) detector functions as a degrader, depth sensor, and stopping target forming an exotic atom. The energy resolution (FWHM) of the characteristic X-rays must be better than approximately 4 keV. The TOF system consists of inner and outer scintillation counters.[23] Each counter is made of a thin (~6 mm thick) and long (~ 180 cm) plastic scintillator. Both ends of the paddles have six silicon photomultipliers (SiPMs). The TOF system provides the trigger and contributes to time of flight and energy deposition measurements. The time resolution obtained experimentally is approximately 0.4 ns.

## 3. Particle Identification

### 3.1 Conventional method

Discrimination between antideuterons and the background requires strict particle identification and background suppression. Conventionally developed particle identification methods rely on reconstructing each event, such as an incident particle, secondary produced particles (pions and protons), and annihilation vertex.[24,25] The performance study used a simulated sample from a Monte Carlo simulation developed by the GAPS collaboration, based on the GEANT4 framework[26] version 10.7. A data set containing approximately $10^4$ channel strips is used to reconstruct incident events (Fig. 4). Each channel provides data on energy assignment, and the TOF counter also provides hit timing data. Hit data are used to reconstruct events and obtain substantial physical quantities for particle identification. In the antideuteron search, antiprotons are the dominant background because only antiparticles can form exotic atoms and create fake antideuteron signals.[14] Baseline studies have demonstrated that a combination of reconstructed quantities could suppress the antiproton background sufficiently.[24] However, developing a solid particle identification method remains challenging due to the complexity of multichannel analysis, the variety of expected signal patterns, and the requirement for high discrimination power.

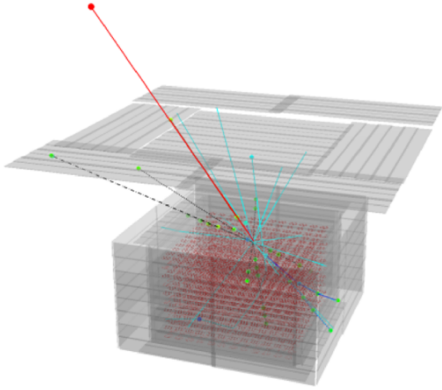

Fig. 4. An event display of an antiparticle incident event example simulated by the GEANT4-based code. The incident particle (red line) is an antideuteron that produces pions through a nuclear annihilation in the instrument.

### 3.2 Machine learning-based method

Considering this background, we investigated a machine learning-based approach using deep neural networks (DNNs) in parallel with the development of the conventional particle identification method.[27] ML has been used in various fields because of its high pattern recognition performance, and it can significantly improve particle identification.

A neural network consists of an input layer, an output layer, and multiple layers between these two layers, as shown in Fig. 5. The edges interconnect the nodes of adjacent layers. Each connection has weights for adjusting the strength of the connection, and each node has a nonlinear activation function in which the output value is calculated by the summation of weighted input values and bias. The output is then fed to the nodes in the next layer. In this way, calculated values transfer to the final output. During iterative learning, the weights of each node update to minimize the loss function, which defines the quantitative difference between estimated and correct answers. As a result, the learning model obtains the preferred output from the given input. This can significantly improve recognition accuracy. In this paper, we refer to a simple model depicted in Fig. 5 as a DNN model.

Because of ML's pattern recognition power, ML-based techniques can provide positive insight into the conventional technique and further improve the antiparticle detection sensitivity of GAPS. For a relatively simple model with limited conditions of incident particle direction and velocity, a DNN model has shown promising results in a previous study.[27] This model used 204 energy depositions from the TOF counters (one from each counter) and 11,520 energy depositions from the Si(Li) detectors (8 strips from 1,440 sensors) as inputs, and the model training was performed to discriminate between antideuterons and antiprotons.[1] However, the direction and velocity ($\beta$) of the simulated incident particles were limited to very narrow ranges ($0.335 < \beta < 0.340$ and $0.250 < \beta < 0.255$ for the monochromatic $\beta$ study). This study evaluates the discrimination performance under more realistic incident conditions and discusses the results of introducing a new ML model to improve the discrimination performance. We discuss exploratory investigations of deep learning approaches for GAPS in the following.

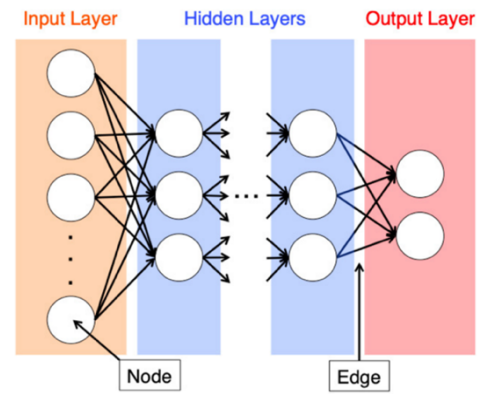

Fig. 5. A conceptual diagram of the deep neural network used in the previous study.[27] The network consists of an input layer, an output layer, and a hidden layer. Circles and arrows indicate the nodes in each layer and the interconnecting edges.

### 4. Machine learning-based Particle Identification

As mentioned in 3.2, we adopted a more realistic simulation in which incident particles are produced isotropically, and the velocity range is expanded to $0.2 < \beta < 0.5$. This is a realistic range of measurable $\beta$ in GAPS and is thus reasonable for high discrimination performance study in a GAPS-like experimental environment. Furthermore, because it is natural to believe that such improved realistic experimental conditions degrade the discrimination performance of the previously used DNN model, we also investigated a new model that combines convolutional neural networks (CNNs) and DNNs.

### 4.1 Convolutional neural networks

A CNN is typically helpful in finding patterns in images to recognize objects, faces, and scenes. In CNNs, filters are applied to inputs of multi-dimensional arrays (e.g., images of two-dimensional pixels). Figure 6 illustrates how each filter of CNN is applied to input arrays (two-dimensional case). Filters have weights in each element and scan the input arrays to extract the features by element-wise product of filter and input array. As a result, a CNN can capture the correlation between pixels.

Because we can regard energy depositions in the GAPS detectors as a spatial data, a three-dimensional (3D) CNN method can be used for our antiparticle identification. Figure 7 shows an overview of the input data flow in our study. A filter of a CNN has a $k \times k \times k$ dimensional structure with weights in each element. The multiplication of each filter by input data results in a set of 3D output arrays, followed by another set of filters in the next convolution layer. Several filters produce output arrays with enhanced specific features, as shown in Fig. 6 (see a line with blocks labeled "3" in a diagonal line). Then, their output is connected to a DNN to make predictions for particle identification.

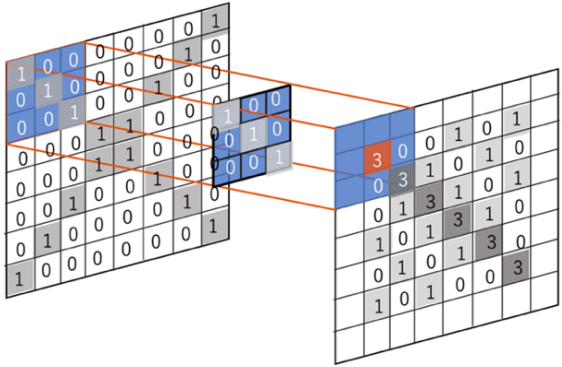

Fig. 6. A conceptual diagram of how CNN's filter is applied to the input data array. In this example, the input data have a two-dimensional array structure, and a filter has a 3 × 3 dimensional structure. The filter captures a diagonal line (top left to bottom right) because of the position of the weights, which returns considerable values only in regions with the same structure as the filter weights.

The previous model in Ref. 27) does not give positional meaning to input data. However, CNN models are trained on the assumption that there are positional correlations in the data. Furthermore, this technique can capture unknown identification features because it uses the spatial distribution of energy depositions, which should have most information available from the detectors.

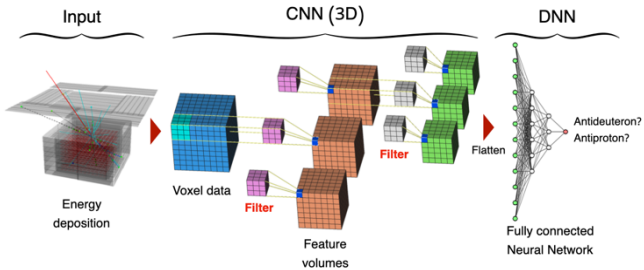

Fig. 7. Overview of CNN application for 3D energy depositions data. Energy depositions are transformed into voxel data and input to CNN. A sequence of filter applications produces 3D data arrays, and the output is then flattened to be connected to a simple DNN for antiparticle identification.

### 4.2 Combination of CNN and physical quantities

Although the CNN technique can improve antiparticle identification performance, it is natural to think that physical quantities (e.g., the number of tracks, hits, and a sum of energy depositions) are expected to improve the performance further. In particular, because the TOF system has only two layers of counters, extracting sufficient features using only the CNN technique is difficult. Introducing TOF-relevant quantities as additional input in such a case should significantly improvement.

To combine these features with CNN outputs, we introduced an additional DNN. First, this DNN takes TOF-relevant quantities as input. Then, the output is merged with the flattened layers of CNN, which calculates the one-dimensional final discriminant value as shown in Fig. 7. Hereafter, we refer to this model as the CNN+DNN model.

### 4.3 Input data

This study merges energy depositions from Si(Li) detectors into 12 × 12 × 10 arrays for CNN input. The eight strips from each Si(Li) wafer are combined so that each element of the 12 × 12 × 10 arrays corresponds to energy depositions from one Si(Li) wafer. As TOF quantities, we used the total energy depositions of outer/inner TOF paddles, the 3D coordinates of the fastest two TOF hits, the numbers of outer/inner TOF paddle hits, and the time-of-flight (11 variables in total). (Note that we can obtain these variables using energy deposition and timing information). The effect of detector resolution is not considered in this study. Furthermore, this study does not consider energy depositions from the characteristic X-rays, and we will revisit them in future studies.

Antideuterons and antiprotons are injected isotropically to make the simulation more realistic in terms of primary particles. The velocity β of the incident antiparticles is produced uniformly with a random distribution and in a range of (1) 0.2 < β < 0.5 and (2) 0.335 < β < 0.340. The latter confirms that the newly developed model outperforms the previous model in the same condition. A data containing 2,000,000 events was constructed to train each antiparticle species. We used 1,600,000 labeled events each as input data for supervised learning. The remaining 400,000 events were used as validation data. This study's data volume is ten times greater than the previous study's.

### 4.4 Training

We used Keras, a Python-based open-source neural network library with the TensorFlow backend, a deep learning framwork.[28] An overview of the studied network structure being studied is shown in Fig. 8. We adopted the rectified linear unit (ReLU) as the activation function. ReLU outputs zero for negative inputs and equal inputs for non-negative inputs. The sigmoid function used in the output layer is defined as

$$\text{sigmoid}(x) = \frac{1}{1+e^{-x}}. \quad (1)$$

Equation (1) is appropriate for binary classification, as in this case. The sigmoid function estimates the class into which input data should be classified by setting the likelihood from 0 to 1.

Table 1 summarizes the primary hyperparameters commonly used in this study. Definitions of each parameter can be found in Ref. 27).

Table 1. Hyperparameters used in this study

| | |
|---|---|
| Batch size | 160 |
| Epoch | 50 |
| Early Stopping | ON |
| Optimizer | Adam[29] |
| Learning rate | 0.00001 |
| Loss function | Binary cross entropy |

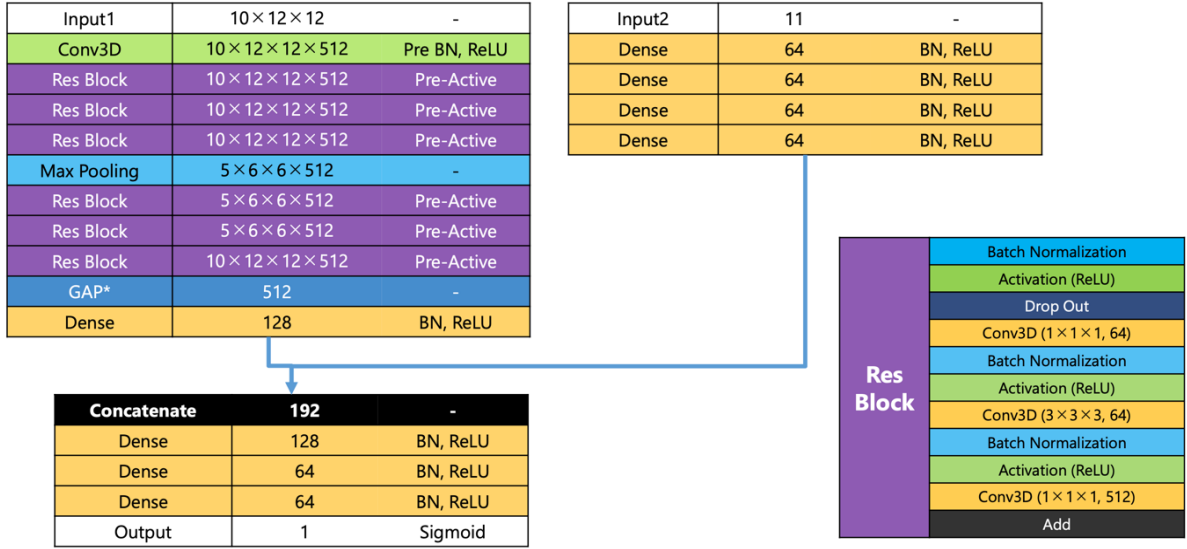

Fig. 8. An overview of the network structure used in this study. "Conv3D" (light green) is a set of three-dimensional CNN filters, and "Dense" is a fully connected layer. The CNN part (top left block) takes energy depositions of $10\times12\times12$ arrays from the Si(Li) detectors, and the DNN part (top right block) takes 11 TOF relevant variables. These outputs are merged in the "Concatenate" section (bottom left block). The detail of "Res Block" (purple) is illustrated in the bottom right block, which has a shortcut connection between input and output to carry out efficient training (input to "Res block" is connected directly to output by "Add" function of Keras).[28]

## 5. Results

### 5.1 Antideuteron and antiproton separation

Figure 9 shows the accuracy profile of the CNN+DNN model for antideuteron and antiproton discrimination training with a velocity range of $0.2 < \beta < 0.5$. The high accuracy achieved in the first few epochs indicates the adequacy of hyperparameter settings. Both the training and validation data suggest that the accuracy improves during iterative learning and approaches 1, showing that overtraining is avoided. The plot shows that the accuracy of the validation data is lower than that of the training data in the last epoch, however, the difference is negligible. The training result for the $0.335 < \beta < 0.340$ case was also confirmed in the same manner.

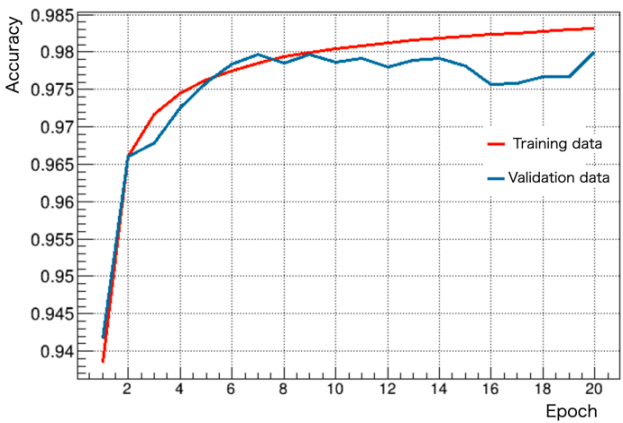

Fig. 9. Learning accuracy of training data (red) and validation data (blue) as functions of the number of epochs. The β range is $0.2 < \beta < 0.5$.

Figure 10 depicts the discriminant distribution of the CNN+DNN model calculated from the validation data with $0.2 < \beta < 0.5$. Most antideuterons (red) and antiprotons (blue) are correctly identified.

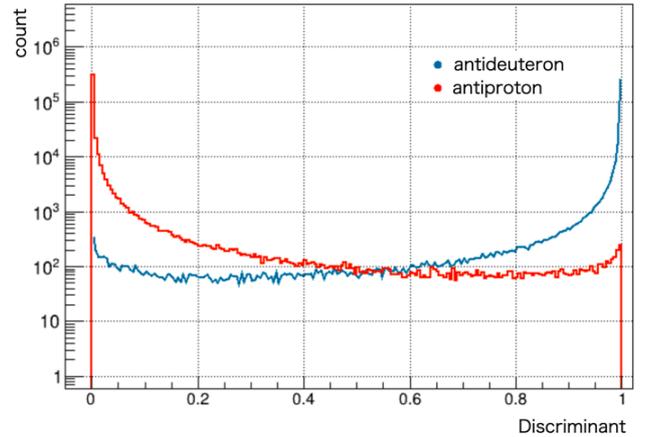

Fig. 10. Output likelihood distributions for validation dataset with $0.2 < \beta < 0.5$. Most antideuteron (blue) and antiproton (red) events are correctly identified.

### 5.2 Receiver operating characteristic curve comparison

Figure 11 compares the previous model (DNN only) and the new model (CNN+DNN) trained with $0.335 < \beta < 0.340$ in terms of the receiver operating characteristic (ROC) curve, where the background rejection power is plotted as a function of signal detection efficiency. Here, a pair of detection efficiency and background rejection power is obtained by selecting events with a discriminant value greater than a signal selection threshold (see Fig. 10). The newly developed model significantly outperforms the previous model by introducing a CNN and TOF-relevant quantities. The new model achieves rejection power greater than $10^4$ with high detection efficiencies.

Figure 12 depicts the ROC curve comparison between (1) and (2) datasets trained using the CNN+DNN model. The wide velocity range reduces the identification performance, as expected. However, it maintains a rejection power of $10^4$ with a signal efficiency lower than 0.75.

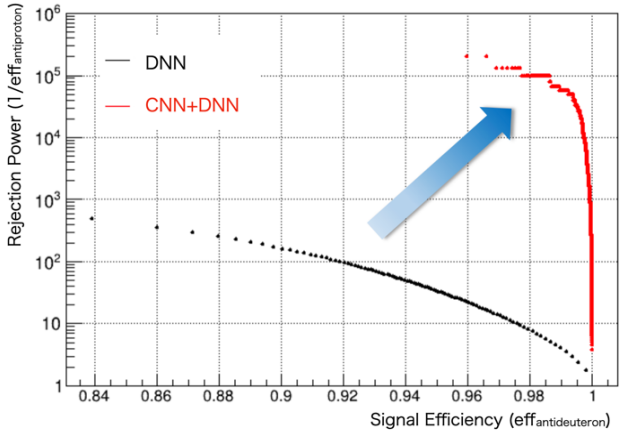

Fig. 11. The ROC curve comparison between the previous model (black) and new (CNN+DNN) model for $0.335 < \beta < 0.340$.

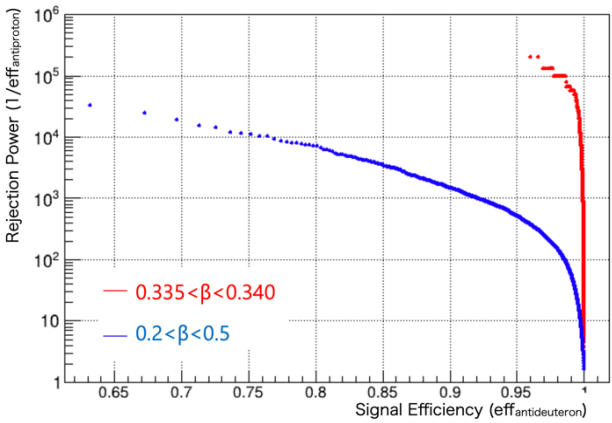

Fig. 12. The ROC curve comparison between datasets of $0.335 < \beta < 0.340$ (red) and $0.2 < \beta < 0.5$ (blue) trained using the CNN+DNN model.

### 5.3 Discussion

This study demonstrates that the newly developed model (CNN+DNN) has potentially high rejection power in a more realistic experimental environment where incident particles enter isotropically and the velocity range is expanded from $0.335 < \beta < 0.340$ to $0.2 < \beta < 0.5$. The findings show the potential of a deep learning approach to study the particle identification capability of the GAPS instrument. This study does not include characteristic X-rays in the simulation data. This implementation will improve the identification accuracy. The effect of detector resolution will be considered in the future studies.

### 6. Conclusion

We investigated ML techniques for GAPS particle identification. We demonstrated that high recognition efficiency could be achieved while removing backgrounds. We intend to conduct experiments under more realistic conditions based on these promising results in future studies. For example, we intend to consider adding detector responses and other particles that could be background events. Furthermore, we intend to extract unknown physical features and feed them to conventional identification methods. The investigation of systematic uncertainty of output value will also be covered.

### Acknowledgments

Funding: This work is supported in the U.S. by NASA APRA grants (NNX17AB44G, NNX17AB46G, and NNX17AB47G), in Japan by JAXA/ISAS Small Science Program FY2017, and in Italy by Istituto Nazionale di Fisica Nucleare (INFN) and the Italian Space Agency through the ASI INFN agreement n. 2018-28-HH.0: "Partecipazione italiana al GAPS - General AntiParticle Spectrometer". M. Yamatani received support from JSPS KAKENHI grant JP22K14065. H. Fuke is supported by JSPS KAKENHI grants (JP17H01136, JP19H05198, and JP22H00147) and Mitsubishi Foundation Research Grant 2019-10038. Y. Shimizu receives support from JSPS KAKENHI grant JP20K04002 and Sumitomo Foundation Grant No. 180322.